%% file: baikalgvd_track_analysis.tex
\documentclass[a4paper,11pt]{article}
\usepackage{pos}
\usepackage{graphicx}

\title{Observations of track-like neutrino events with Baikal-GVD}

\input{baikalgvd_author_list_icrc2021.tex}

\abstract{
The Baikal Gigaton Volume Detector (Baikal-GVD) is a km$^3$-scale neutrino detector currently under construction in Lake Baikal, Russia.
The detector currently consists of 2304 optical modules arranged on 64 vertical strings.
Further extension of the array is planned for March 2022.
The data from the partially complete array have been analyzed using a $\chi^2$-based track reconstruction algorithm.
After suppression of the downward-going atmospheric muon background, a flux of upward-going neutrino events is observed, dominated by the atmospheric neutrinos.
The observed flux is in good agreement with Monte Carlo predictions.
}

\FullConference{%
  *** 37th International Cosmic Ray Conference (ICRC2021), ***\\
  *** 12-23 July 2021 ***\\
  *** Berlin, Germany - Online ***
}

\begin{document}
\maketitle


\section{Introduction}

\begin{figure}
  \centering
  \includegraphics[height=8.0cm]{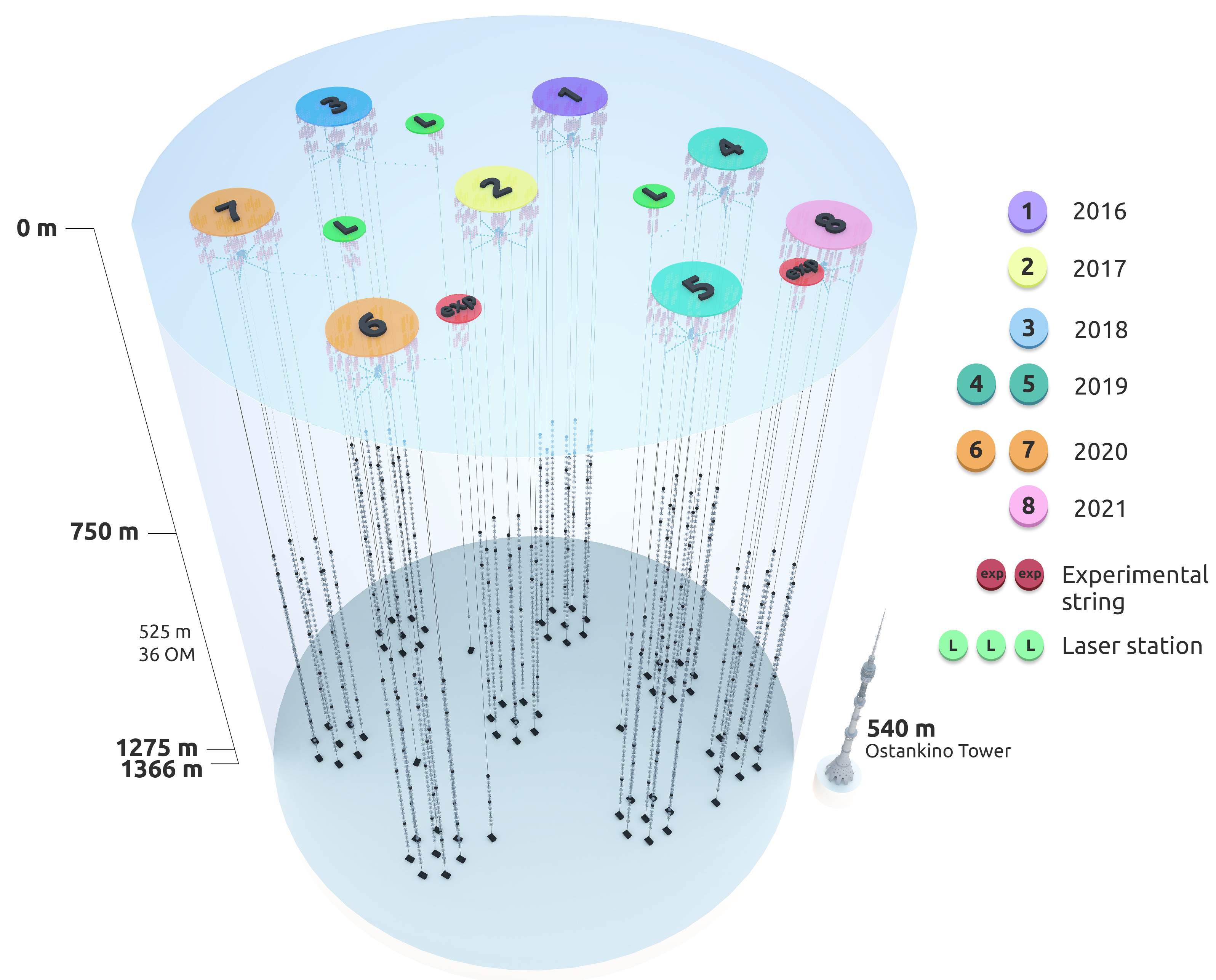}
  \includegraphics[height=8.0cm]{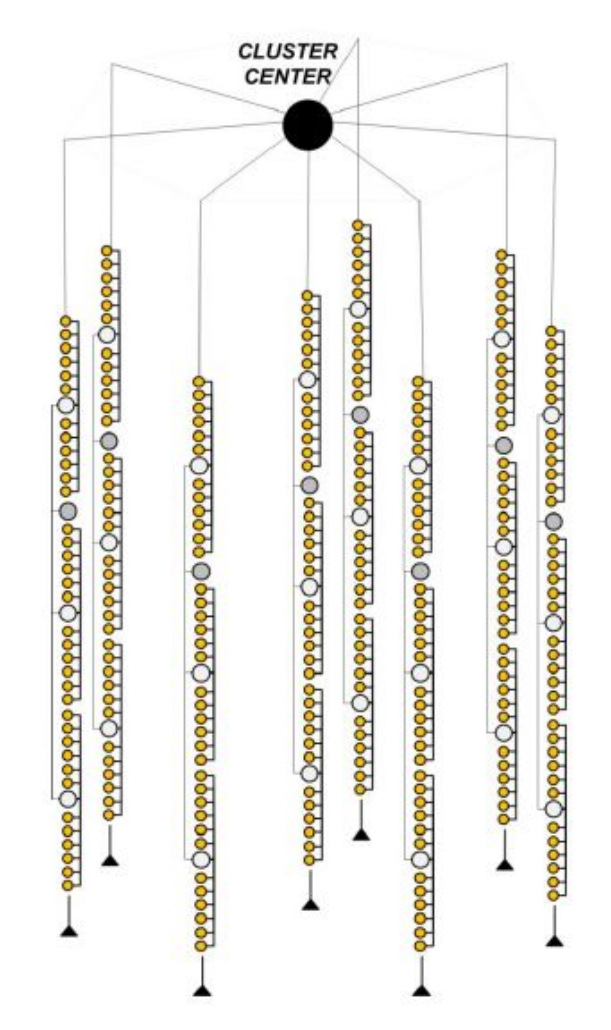}
  \caption{Left: Schematic view of the Baikal-GVD detector.
           The yearly progression of the detector deployment is shown in the legend.
           Right: The Baikal-GVD cluster layout (vertical scale compressed).}
  \label{fig:baikal_gvd}
\end{figure}

The Baikal Gigaton Volume Detector (Baikal-GVD) is a cubic-kilometer scale underwater neutrino detector currently under construction in Lake Baikal (Russia) at
a depth of 1366~m \cite{Baikal-GVD}.
The detector uses 10-inch photomultiplier tubes (PMTs) to detect the Cherenkov light
from charged particles produced in a neutrino interaction.
The detector elements are arranged along vertical strings which are in turn arranged in heptagonal clusters.
Each cluster is equipped with 288 PMTs arranged on 8 strings.
Eight clusters, with a total of 2304 PMTs, have already been deployed (see Fig.~\ref{fig:baikal_gvd}).
Six more clusters are scheduled for deployment in the next three years.

Events resulting from charged current (CC) interactions of muon (anti-)neutrinos, as well as $\tau$-neutrino CC interactions followed by leptonic $\tau$ decays, will have a track-like topology in Baikal-GVD.
A fast $\chi^2$-based reconstruction algorithm has been developed to reconstruct such track-like events \cite{Baikal_track_paper,Baikal_track_reco_icrc2021}.
A simple cut-based analysis has been employed to select neutrino events coming from below horizon and suppress the background of misreconstructed atmospheric muon events.
The analysis is performed in two different modes: a single-cluster analysis, where data from each detector cluster are treated independently, and a multi-cluster analysis, where data from all clusters are joined together and are analyzed.
For both the analysis types, the analysis cuts have been optimized for the detection of atmospheric neutrinos.
The reconstruction algorithm has been applied to data collected in 2019 with the first five operational clusters of Baikal-GVD, yielding successful observations of upgoing atmospheric neutrinos in the single-cluster analysis.
The mutli-cluster analysis of the experimental dataset is currently pending a final decision on the data calibration readiness.

The paper is organized as follows. 
The dataset and Monte Carlo simulations are briefly described in Section~\ref{sect:dataset}.
The track reconstruction algorithm is introduced in Section \ref{sect:reco}.
The single-cluster and multi-cluster analyses are presented in Sections~\ref{sect:analysis} and \ref{sect:mutli_cluster}, respectively.
Section~\ref{sect:conclusion} concludes the paper.

\section{Experimental dataset and Monte Carlo simulations}
\label{sect:dataset}
In this work we use a dataset collected from the first five operational clusters of Baikal-GVD between April 1 and June 30, 2019.
This period is characterized by relatively quiet optical noise levels
(see \cite{baikal_optical_noise} for a review of the optical noise conditions at the Baikal-GVD site).
The average measured rate of noise hits observed by the OMs in this period ranges from $\approx 15$~kHz (at the bottommost detector storeys) to $\approx 50$~kHz (at the topmost storeys).
The total single-cluster-equivalent livetime of the dataset is $\approx 323$~days.
The livetime and number of operating optical modules for each cluster are summarized in Table~\ref{table:dataset}.
The analysis livetime for two-cluster events corresponds to $\approx 2$~months of the 5-cluster detector operation.

\begin{table}[h!]
  \begin{center}
    \begin{tabular}{|c|c|c|}
      \hline
      \textbf{GVD cluster} & \textbf{Number of active OMs} & \textbf{Dataset duration, days}\\
      \hline
      1 & 270 & 68\\
      2 & 273 & 72\\
      3 & 288 & 74\\
      4 & 288 & 61\\
      5 & 288 & 47\\
      \hline
      1--5 combined single-cluster & 1407 & 323\\
      \hline
    \end{tabular}
    \caption{Basic characteristics of the dataset used in this work
(April 1 -- June 30, 2019).
}
    \label{table:dataset}
  \end{center}
\end{table}

A dedicated Monte Carlo (MC) simulation has been performed in order to estimate the expected event rates and optimize the analysis at Baikal-GVD.
The present study is focused on reconstruction of the track-like events, therefore, only CC interaction of muon (anti-) neutrino is considered.
A simple neutrino generator code was used to simulate CC interactions of muon neutrino and muon anti-neutrino with nuclei in water.
The simulated neutrino energy spectrum and angular distribution follow the Bartol flux model \cite{Bartol1996}.
The neutrino-nuclei CC interaction cross sections were calculated using the CTEQ4M parton distribution functions \cite{CTEQ4M}.
The simulation covers the neutrino energy range from 10 GeV up to 100 TeV.

For the simulation of atmospheric muon bundles, CORSIKA~5.7 \cite{CORSIKA} has been used 
to model the Cosmic Ray (CR) interactions in the atmosphere and to propagate the secondary particles down to the lake surface level.
The simulation incorporates the QGSJET model of hadronic interactions \cite{QGSJET}.
For the CR composition, we use a mutli-component model based on KASCADE data \cite{Wiebel_Sooth_1999}.
The simulation covers the energy range of the primary CR nuclei from 240 GeV up to 2 PeV.
The propagation of the muons from the lake surface down to and through the detector volume is handled with MUM \cite{MUM}.
Further details on the track-like event simulation chain can be found in \cite{Baikal_track_paper}.

\section{Data analysis procedure}
\label{sect:reco}
After applying the necessary calibrations \cite{Baikal_calibration,Baikal_time_calibration}, the experimental dataset is passed to a $\chi^2$-based track reconstruction algorithm,
which includes a hit selection procedure, a vector-sum prefit, and a $\chi^2$-like track fit.
The algorithm takes as input a list of PMT hits recorded in a 5~$\mu$s time window.
The hit information includes the hit time and charge.
The reconstruction algorithm takes into account the PMT coordinates as determined with the acoustic positioning system \cite{Baikal_positioning}.
Each event is assumed to have only one muon track.
For a full description of the algorithm see \cite{Baikal_track_paper}.
Further development and variations of the algorithm are discussed in \cite{Baikal_track_reco_icrc2021}.

\section{Single-cluster analysis}
\label{sect:analysis}

The single-cluster dataset, described in Section~\ref{sect:dataset}, has been 
processed with the $\chi^2$-based track reconstruction algorithm, as discussed in Section~\ref{sect:reco}.
A minimal requirement of at least 8 hit OMs on at least two detector strings has been applied to ensure favorable conditions for accurate reconstruction of the zenith angle.
Additionally, a good fit convergence status is required.
This yields 9.8 million reconstructed events for the combined dataset from the 5 detector clusters.
The same reconstruction and analysis procedure has been applied to the simulated atmospheric muon and neutrino samples.

\begin{figure}
  \centering
  \includegraphics[height=7.5cm]{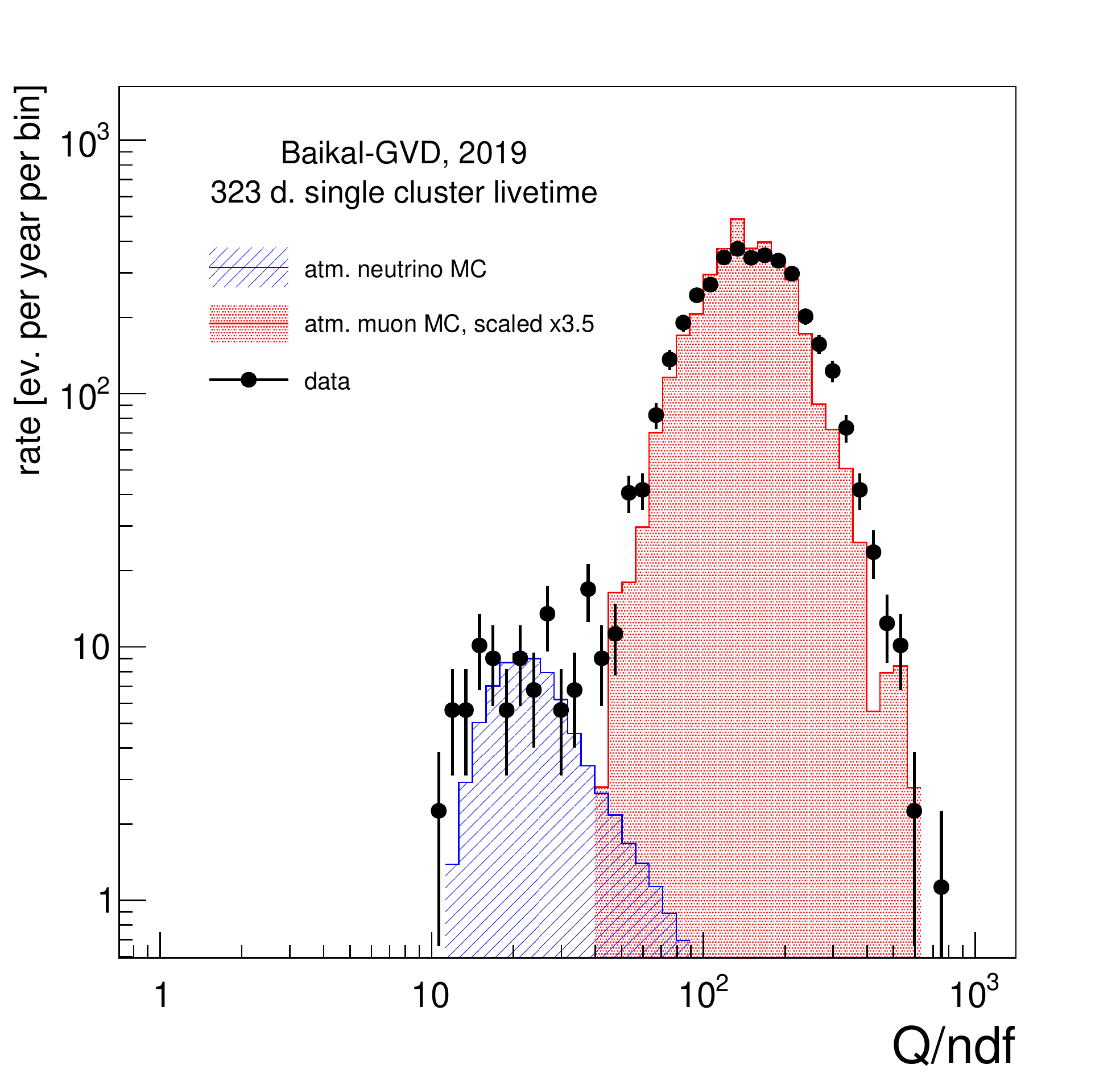}
  \caption{Distribution of the fit quality parameter for tracks, reconstructed as upward-going with cos $\theta<-0.5$ in the single-cluster analysis. 
The Baikal-GVD data are shown by black points with statistical error bars.
The MC predictions for atmospheric muons (scaled by a factor 3.5) and atmospheric neutrinos (not scaled) are shown in red and blue, respectively.
}
  \label{fig:track_analysis_results}
\end{figure}

The majority of the atmospheric muons can be suppressed by a cut on the zenith angle.
The remaining background of misreconstructed events can be further suppressed by tight cuts based on the fit quality and on other parameters.
Figure~\ref{fig:track_analysis_results} shows the distribution of the fit quality $Q$ divided by the number of degrees of freedom (ndf) for events reconstructed with $\theta>120^\circ$ (cos $\theta<-0.5$). 
For this figure we have re-scaled the atmospheric muon MC by a factor of 3.5 (see \cite{Baikal_track_paper} for details).
The data are shown in black.
The MC expectations for atmospheric muons and upgoing atmospheric neutrinos are shown in red and blue, respectively.
As one can see, the fit quality variable allows an efficient separation of the upgoing track-like neutrino events from the background of atmospheric muon events, misreconstructed as upgoing. 
The shape of the fit quality distribution observed for experimental data is compatible with the MC prediction.

\begin{figure}
  \centering
  \includegraphics[width=0.49\textwidth]{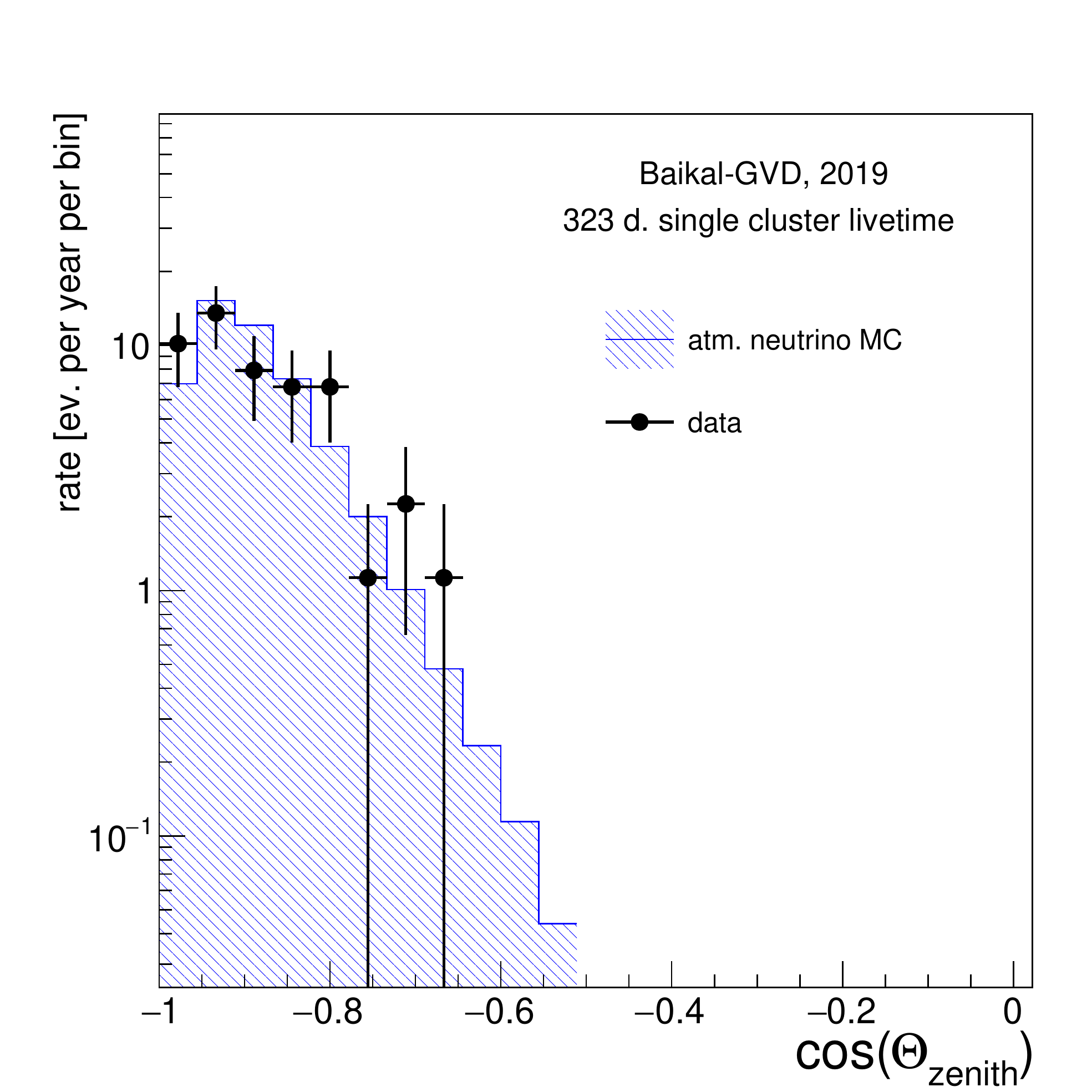}
  \includegraphics[width=0.49\textwidth]{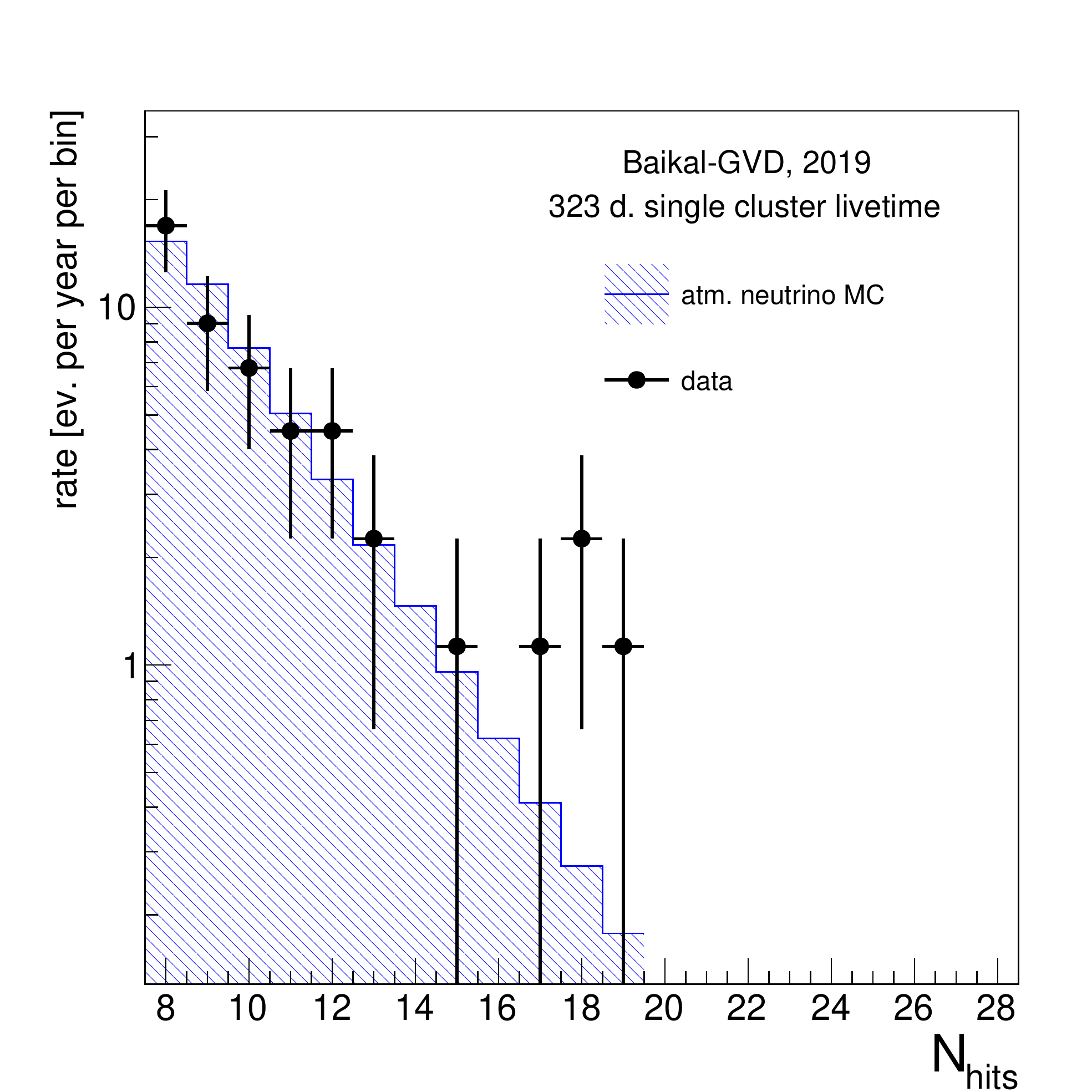}
  \caption{Results of the search for upgoing track-like neutrino events in the single-cluster dataset (323 days single-cluster equivalent livetime).
           Left: zenith angle distribution.
           Right: distribution of the number of hits used in reconstruction.
           The black points and the blue filled area show the Baikal-GVD data and the MC prediction for atmospheric neutrinos, respectively.
           The expected number of background atmospheric muon events after the quality cuts is $\lesssim$1 and is therefore not shown.
 }
  \label{fig:track_analysis_results_nu}
\end{figure}

After a cut on the fit quality parameter ($Q/ndf < 32$) we suppress the remaining background due to atmospheric muons stepwise by cuts on other parameters. 
This includes the visible track length ($L > 42$~m), average distance from the track to the hit OMs ($\rho<18$~m), combined charge of the observed hits ($C > 18$~p.e.), estimated zenith angle error ($\theta_{err}<2^\circ$), as well as cuts on the width of the time residual distribution and additional hit likelihood variables.
Using these analysis cuts, a total of 44 neutrino candidate events were found in the experimental data while the expectation from the atmospheric neutrino MC simulation is $43.6$ $\pm \, 6.6$ (stat.) events, and the expectation for the atmospheric muon background is $\lesssim$1 event.
The median energy of the neutrino events, according to the MC simulations, is $\approx 500$ GeV.
The resulting zenith angle distribution of the neutrino candidate events is shown in left panel of Fig.~\ref{fig:track_analysis_results_nu}.
The distribution of the number of hits is shown in right panel of Fig.~\ref{fig:track_analysis_results_nu}.
A good agreement with the MC prediction can be noted.

\begin{figure}
  \centering
  \includegraphics[height=7.5cm]{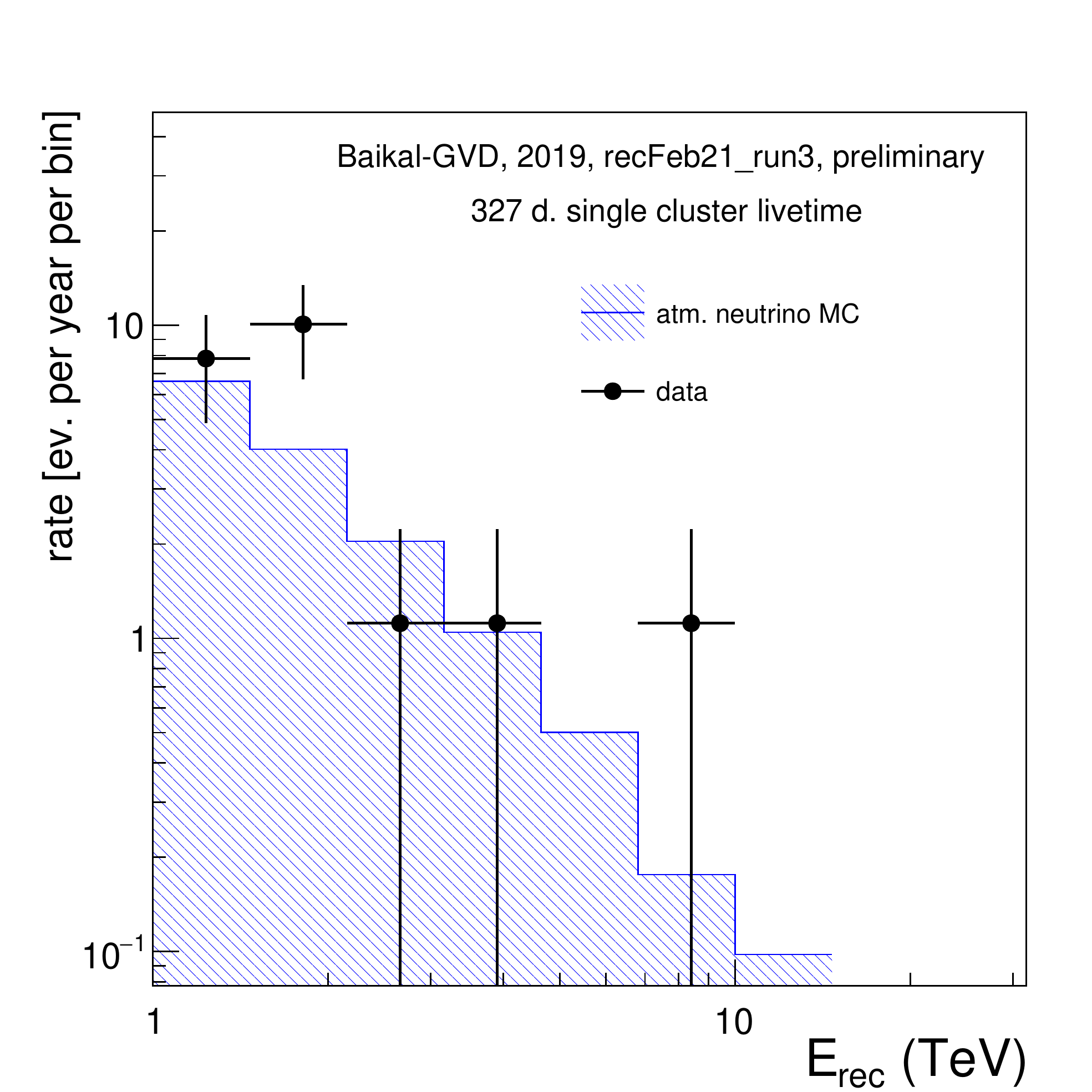}
  \caption{Reconstructed energy distribution for a set of neutrino candidate events.
}
  \label{fig:track_analysis_results_energy}
\end{figure}

An energy estimator algorithm has been developed to estimate the visible energy of track-like neutrino events based on the brightness associated with the track \cite{Baikal_track_reco_icrc2021}.
This method is applicable for neutrino energies higher than 1 TeV.
An example of the measured energy distribution for a set of neutrino events is shown in Fig.~\ref{fig:track_analysis_results_energy}.
As one can see, the observed distribution (black points) is in reasonable agreement with the MC prediction for atmospheric neutrinos (blue histogram).
Note that the analysis reported here -- by construction -- disfavours the highest energy neutrino events.
A dedicated analysis optimized for high energy ($E > 10$~TeV) is in preparation.

\section{Multi-cluster analysis}
\label{sect:mutli_cluster}

The multi-cluster analysis uses the same reconstruction and analysis procedure as the single-cluster analysis, with the only difference that this time the event is required to include data from at least two clusters.
This requirement reduces the rate of events to $\approx$ 10\% of the single-cluster event rate.
The neutrino event selection is analogous to the selection used in the single cluster analysis, with the following differences:
\begin{itemize}
  \item the zenith angle cut is set to $\theta>90^\circ$;
  \item the minimal track length is set to 250 m;
  \item the fit quality cut is set to $Q/ndf <50$;
  \item the estimated zenith angle error cut is set to $\theta_{err}<0.5^\circ$;
  \item there is no cut on the average track-hit distance.
\end{itemize}

\begin{figure}
  \centering
  \includegraphics[height=7.5cm]{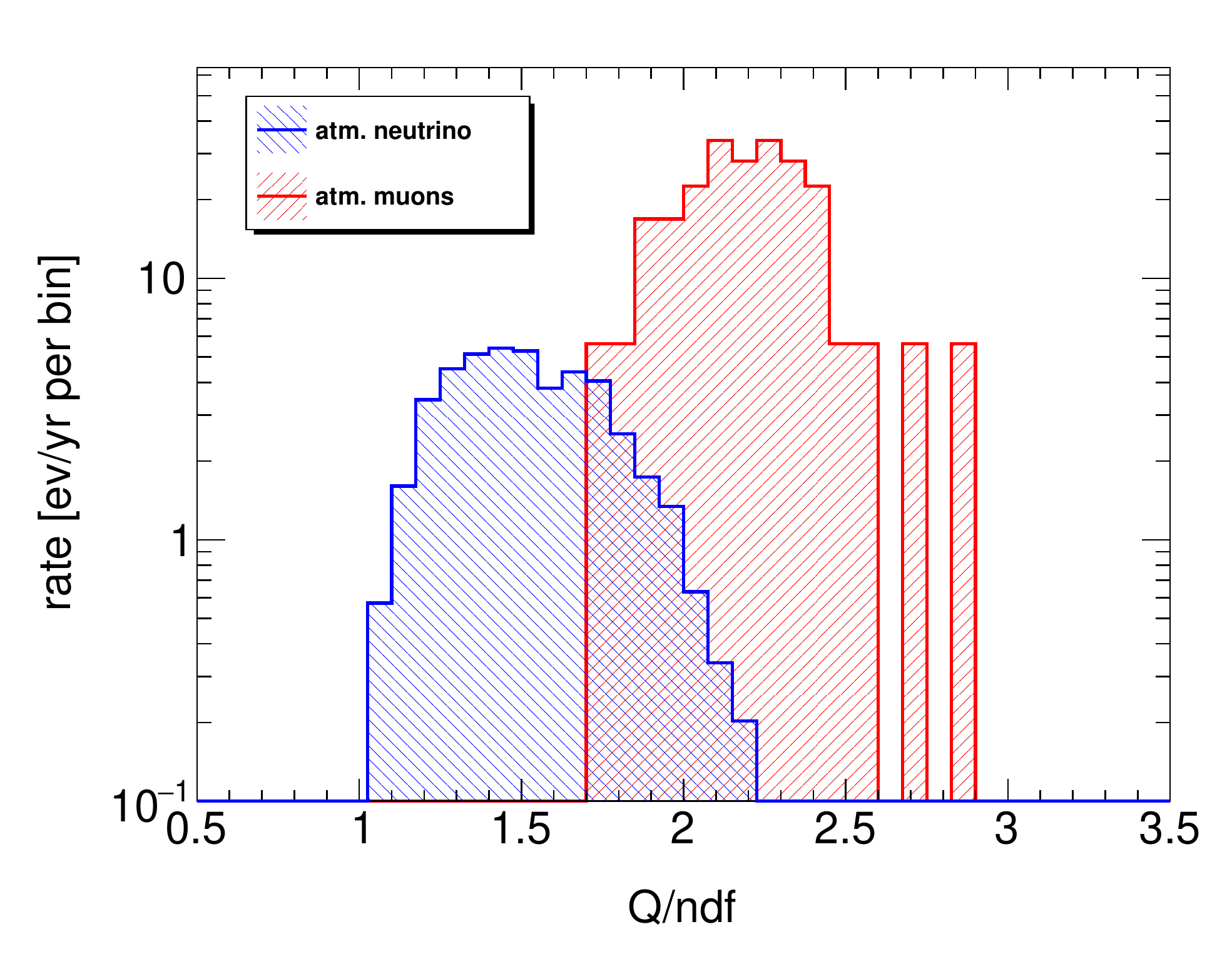}
  \caption{Distribution of the fit quality parameter for tracks reconstructed as upward-going with cos $\theta<0$ and track length $L > 250$~m in the multi-cluster analysis. 
The MC predictions for atmospheric muons and atmospheric neutrinos are shown in red and blue, respectively.
}
  \label{fig:track_analysis_results_mutlicluster}
\end{figure}

The fit quality distribution after the cuts on the zenith angle and track length is shown in Fig.~\ref{fig:track_analysis_results_mutlicluster} for both the atmospheric muon MC and the atmospheric neutrino MC. 
It needs to be noted that the equivalent livetime of the atmospheric muon MC shown here is only 65 days.
With all of the analysis cuts applied the expected rate of atmospheric neutrinos is 29.4 events per year, which corresponds to $\approx$ 4.9 events for a two-month dataset.
The median neutrino energy for the atmospheric neutrino events in this analysis is $\approx$~4~TeV.
The analysis efficiency can be improved by a factor of 2 or more by employing more advanced reconstruction and event selection algorithms \cite{Baikal_track_reco_icrc2021}.

\section{Conclusion}
\label{sect:conclusion}
Baikal-GVD is a new TeV-PeV neutrino telescope under construction in Lake Baikal, Russia.
The telescope has been collecting data in partial configurations since 2016.
As of April 2021, the detector consists of 8 clusters (64 strings, 2304 optical modules) with a total effective volume of 0.4 km$^3$.
A $\chi^2$-based track reconstruction algorithm has been developed to reconstruct track-like events in Baikal-GVD.
The algorithm has been applied to a combined dataset of single-cluster events collected from the first five operational clusters of Baikal-GVD over a two month period in spring 2019, yielding 44 neutrino candidate events.
The observed flux and energy distribution of the neutrino events is in good agreement with Monte Carlo predictions.
A mutli-cluster analysis using the same reconstruction algorithm has also been prepared, currently pending a final decision on the multi-cluster calibration readiness.
Further analysis developments promise to improve the neutrino detection efficiency, angular resolution, and energy measurements.

\section*{Acknowledgements}
The work was partially supported by RFBR grant 20-02-00400.
The CTU group acknowledges the support by European Regional Development Fund-Project No. CZ.02.1.01/0.0/0.0/16\_019/0000766. 
We also acknowledge the technical support of JINR staff for the computing facilities (JINR cloud).

\end{document}

%% file: baikalgvd_author_list_icrc2021.tex
\author[a]{V.A.~Allakhverdyan}
\author[b]{A.D.~Avrorin}
\author[b]{A.V.~Avrorin}
\author[b]{V.M.~Aynutdinov}
\author[c]{R.~Bannasch}
\author[d]{Z.~Barda\v{c}ov\'{a}}
\author[a]{I.A.~Belolaptikov}
\author[a]{I.V.~Borina}
\author[a,1]{V.B.~Brudanin}
\author[e]{N.M.~Budnev}
\author[a]{V.Y.~Dik}
\author[b]{G.V.~Domogatsky}
\author[b]{A.A.~Doroshenko}
\author[a,d]{R.~Dvornick\'{y}}
\author[e]{A.N.~Dyachok}
\author[b]{Zh.-A.M.~Dzhilkibaev}
\author[d]{E.~Eckerov\'{a}}
\author[a]{T.V.~Elzhov}
\author[f]{L.~Fajt}
\author[g,1]{S.V.~Fialkovski}
\author[e]{A.R.~Gafarov}
\author[b]{K.V.~Golubkov}
\author[a]{N.S.~Gorshkov}
\author[e]{T.I.~Gress}
\author[a]{M.S.~Katulin}
\author[c]{K.G.~Kebkal}
\author[c]{O.G.~Kebkal}
\author[a]{E.V.~Khramov}
\author[a]{M.M.~Kolbin}
\author[a]{K.V.~Konischev}
\author[h]{K.A.~Kopa\'{n}ski}
\author[a]{A.V.~Korobchenko}
\author[b]{A.P.~Koshechkin}
\author[i]{V.A.~Kozhin}
\author[a]{M.V.~Kruglov}
\author[b]{M.K.~Kryukov}
\author[g]{V.F.~Kulepov}
\author[h]{Pa.~Malecki}
\author[a]{Y.M.~Malyshkin}
\author[b]{M.B.~Milenin}
\author[e]{R.R.~Mirgazov}
\author[a]{D.V.~Naumov}
\author[a]{V.~Nazari}
\author[h]{W.~Noga}
\author[b]{D.P.~Petukhov}
\author[a]{E.N.~Pliskovsky}
\author[j]{M.I.~Rozanov}
\author[a]{V.D.~Rushay}
\author[e]{E.V.~Ryabov}
\author[b]{G.B.~Safronov}
\author[a]{B.A.~Shaybonov}
\author[b]{M.D.~Shelepov}
\author[a,d,f]{F.~\v{S}imkovic}
\author[a]{A.E. Sirenko}
\author[i]{A.V.~Skurikhin}
\author[a]{A.G.~Solovjev}
\author[a]{M.N.~Sorokovikov}
\author[f]{I.~\v{S}tekl}
\author[b]{A.P.~Stromakov}
\author[a]{E.O.~Sushenok}
\author[b]{O.V.~Suvorova}
\author[e]{V.A.~Tabolenko}
\author[e]{B.A.~Tarashansky}
\author[a]{Y.V.~Yablokova}
\author[c]{S.A.~Yakovlev}
\author*[b]{D.N.~Zaborov}

\affiliation[a]{Joint Institute for Nuclear Research, Dubna, Russia}
\affiliation[b]{Institute for Nuclear Research, Russian Academy of Sciences, Moscow, Russia}
\affiliation[c]{EvoLogics GmbH, Berlin, Germany}
\affiliation[d]{Comenius University, Bratislava, Slovakia}
\affiliation[e]{Irkutsk State University, Irkutsk, Russia}
\affiliation[f]{Czech Technical University in Prague, Prague, Czech Republic}
\affiliation[g]{Nizhny Novgorod State Technical University, Nizhny Novgorod, Russia}
\affiliation[h]{Institute of Nuclear Physics of Polish Academy of Sciences (IFJ~PAN), Krak\'{o}w, Poland}
\affiliation[i]{Skobeltsyn Institute of Nuclear Physics, Moscow State University, Moscow, Russia}
\affiliation[j]{St.~Petersburg State Marine Technical University, St.Petersburg, Russia}

\note{Deceased.}

\emailAdd{zaborov@inr.ru}

\renewcommand{\printHeadAuthors}{D. Zaborov for Baikal-GVD}